\documentclass[letterpaper,aps,pra,twocolumn,english,showpacs,superscriptaddress]{revtex4-1}

\usepackage{tikz}
\usepackage{pslatex}
\usepackage[T1]{fontenc}
\usepackage[latin1]{inputenc}
\usepackage{amssymb, amsmath, amsfonts, bm}
\usetikzlibrary{decorations.pathmorphing}
\usetikzlibrary{decorations.pathreplacing}
\usepackage{natbib}
\providecommand{\LyX}{L\kern-.1667em\lower.25em\hbox{Y}\kern-.125emX\@}

\newcommand{\Eref}[1]{Eq.~(\ref{#1})}
\definecolor{dgreen}{RGB}{50,128,128}
\usepackage{multirow}
\usepackage{graphicx}
\input{epsf}
\usepackage{babel}

\begin{document}
\title{$p$-Wave Optical Feshbach Resonances in ${}^{171}$Yb}

\author{Krittika Goyal}
\altaffiliation{Krittika Goyal previously published under the name Krittika Kanjilal}
\affiliation{Center for Quantum Information and Control (CQuIC), University of New Mexico, Albuquerque NM 87131}
\affiliation{Department of Physics and Astronomy, University of New Mexico, Albuquerque NM 87131}

\author{Iris Reichenbach}
\affiliation{Max Planck Institute for the Physics of Complex Systems, N\"{o}thnitzer Str. 38, D-01187 Dresden, Germany}

\author{Ivan Deutsch}
\affiliation{Center for Quantum Information and Control (CQuIC), University of New Mexico, Albuquerque NM 87131}
\affiliation{Department of Physics and Astronomy, University of New Mexico, Albuquerque NM 87131}

\begin{abstract}
We study the use of an optical Feshbach resonance to modify the $p$-wave interaction between ultracold polarized ${}^{171}$Yb spin-1/2 fermions.  A laser exciting two colliding atoms to the $^1S_0 + {}^3P_1$ channel can be detuned near a purely-long-range excited molecular bound state.  Such an exotic molecule has an inner turning point far from the chemical binding region and thus three-body-recombination in the Feshbach resonance will be highly suppressed in contrast to that typically seen in a ground state $p$-wave magnetic Feshbach resonance.  We calculate the excited molecular bound-state spectrum using a multichannel integration of the Schr\"{o}dinger equation, including an external perturbation by a magnetic field.  From the multichannel wave functions, we calculate the Feshbach resonance properties, including the modification of the elastic $p$-wave scattering volume and inelastic spontaneous scattering rate.  The use of magnetic fields and selection rules for polarized light yields a highly controllable system.  We apply this control to propose a toy model for three-color superfluidity in an optical lattice for spin-polarized ${}^{171}$Yb, where the three colors correspond to the three spatial orbitals of the first excited $p$-band.  We calculate the conditions under which tunneling and on-site interactions are comparable, at which point quantum critical behavior is possible.
\end{abstract}
%\pacs{34.50.-s,34.10.+x}
\maketitle

\section{Introduction}

Alkaline-earth-like atoms  are of increasing interest for applications in quantum control, including optical atom clocks~\cite{diddams2003}, quantum computing~\cite{derevianko2004,hayes2007,daley2008,gorshkov2009}, and simulations of condensed matter systems~\cite{gorshkov2010}.  Experimental advances are proceeding at a steady pace with demonstrations of a variety of important milestones, including clocks that now surpass the cesium standard~\cite{ludlow2008}, Bose-Einstein condensation in isotopes of ytterbium~\cite{takasu2003}, calcium~\cite{kraft2009}, and strontium~\cite{stellmer2009,escobar2009}, Fermi degenerate gases~\cite{fukuhara2007}, and the superfluid-to-mott-insulator quantum phase transition~\cite{ybmott}.  

Another important ingredient in the quantum-control toolbox is the ability to control the interatomic interactions.  Feshbach resonances have played an essential role in such manipulation of alkali-metal degenerate gases, allowing for the observation of the BEC-BCS crossover~\cite{greiner2003}.  Whereas in alkali gases Feshbach resonances can be induced via magnetic fields that couple different channels in the electronic ground state, in alkaline-earth-like atoms this is not possible because of the lack of hyperfine structure in the ground $^1S_0$ state.  An alternative is to employ an optical Feshbach resonance (OFR) by laser-coupling two scattering ground-state atoms to a meta-stable bound molecule in an excited-state potential~\cite{bohnandjulienne97}.  Alkaline earths are particularly well suited to OFRs due to the existence of narrow intercombination lines of the kind studied for optical clocks, $^1S_0 \rightarrow {}^3P_J$~\cite{ciurylo2005}.   Photoassociation spectroscopy has been used to measure narrow molecular resonances in the  $^1S_0 + {}^3P_1$ channel~\cite{zelevinsky2006,enomoto2008}, an important first step toward implementation of OFRs.  

In previous work we studied the use of OFRs to manipulate nuclear spin coherence in fermionic, spin-1/2, $^{171}$Yb using OFRs associated with s-wave collisions~\cite{iris_2009}.  In the work presented here, we extend our study to $p$-wave OFRs of this species.  The ability to manipulate $p$-wave collisions could open the door to studies of nonconventional superfluidity and other exotic quantum phases of matter~\cite{anisopsf}.  Prior observations of $p$-wave magnetic Feshbach resonances in alkalis proved to be too lossy for quantum coherent control~\cite{pwaveMFRJin}.  Inelastic collisions are believed to be enhanced in these resonances because the $p$-wave scattering states are well localized behind the centrifugal barrier~\cite{pwaveMFRJin}.  They thus have a very large Franck-Condon overlap with more tightly bound molecules below the Feshbach threshold, which leads to exothermic transitions.  The use of OFRs can potentially mitigate this effect.  In particular, purely-long-range (PLR) molecular states existing in excited state potentials can be coupled optically to ground-state $p$-wave channels \cite{enomoto2008}.  These PLR states, arising from avoided crossings in the excited state hyperfine structure, have inner turning points at $\sim 50 a_0$, and are thus well separated from the chemical binding region.  Inelastic collisions to bound-ground molecules via excitation to PLR states should be highly suppressed.  In this case, heating due to spontaneous emission will be the dominant source of inelastic collisions, but this too can be suppressed through off-resonance excitation.

In addition to suppressing inelastic recombination, OFRs offer opportunities for quantum control beyond what is possible with magnetic Feshbach resonances.  For example, in $p$-wave collisions the projection of rotational angular momentum along a given axis is a new degree of freedom that can affect the symmetry of the order parameter in $p$-wave superfluidity~\cite{pxpy}.  In the presence of a bias magnetic field and for appropriate choices of laser polarization, we can address these degrees of freedom and control the scattering length associated with different projection quantum numbers. If an optical lattice trapping potential is added, a variety of rich phenomena can be explored with such control.  For example, the three projections of angular momentum translate into three orbitals of a $p$-band in the first excited vibrational state of an optical lattice~\cite{scarola2005,pxpyhubbard}. With control of $p$-wave collisions of spin-polarized fermions, one can obtain a Hubbard model similar to the one that gives rise to 3-color superfluidity and trionic phases and is an important model of QCD~\cite{rapp2007}.

In this article we study the use of an OFR to control $p$-wave collisions in $^{171}$Yb by exciting near photoassociation resonances of the $^1S_0 + {}^3P_1$ channel.  After reviewing the system and the formalism for calculating the optically controlled scattering properties, we calculate the energy spectrum and scattering lengths including the presence of a magnetic field, which allows for polarization-dependent control of the interaction.  We apply this to a toy model of 3-color superfluidity to give a benchmark of the performance of the $p$-wave OFR and summarize our results.  

\section{$p$-wave Photoassociation Resonances}
We consider spin-polarized $^{171}$Yb, with nuclear spin $i = 1/2$, for which $s$-wave collisions are forbidden and $p$-waves dominate at low temperature. The essential formalism for describing the system, in the absence of an external magnetic field, was given in \cite{iris_2009}.  We review the salient points here.  The two-atom states in each of the collision channels are governed by an effective potential of the form
\begin{equation}
V_{\text{eff}} = \frac{R(R+1)}{2\mu r^2}+V_{\text{BO}}(r)+V_{\text{HF}}+V_{\text{mag}},
\label{Veff}
\end{equation}
where $V_{\text{BO}}$ is the Born-Oppenheimer potential in Hund's case-(c), $V_{\text{HF}}$  is the hyperfine interaction, and $V_{\text{mag}}$ 
is the interaction with external magnetic fields.  Here and throughout we set $\hbar=1$ and we use atomic units.  For the ground $^1S_0+ {}^1S_0$ collision there is only one channel, the nuclear spin triplet state $I=1, m_I=1$. There is no hyperfine interaction and we neglect the very small magnetic interaction with the nuclear magneton.  As we are interested only in the near-threshold scattering states of this channel, the ground Born-Oppenheimer potentials can be approximated in a modified Leonard-Jones form~\cite{improved_pot},
\begin{equation}\label{gnd_pot}
 V_{\text{BO}}^{(g)}(r) = \frac{C_{12}^{(g)}}{r^{12}}-\frac{C_6^{(g)}}{r^6}-\frac{C_8^{(g)}}{r^8},
 \end{equation}
where $C_6^{(g)}=1931.7 \rm{a.u.}$, $C_8^{(g)}=1.93 \times 10^5 \rm{a.u.}$, and $C_{12}^{(g)}= 1.03409\times10^9 \rm{a.u.}$ ~\cite{improved_pot}. Since we are considering $p$-wave scattering, the rotational angular momentum is $R=1$. The system is not prepared in a state with a fixed projection of $R$, and thus the atoms can scatter with any allowed value of $m_R=-1, 0, 1$ relative to a space-fixed quantization axis, defined by the magnetic field.  We obtain the scattering wave functions corresponding to the above potential numerically, using the Numerov method for integration~\cite{numerov1, numerov2}.

In the excited $^1S_0+{}^3P_1$ channel, the description is more complicated. The electronic Born-Oppenheimer potentials are taken in the Hund's case-(c),
\begin{equation}
 \label{HBO}
 V_{\text{BO}}^{(e)}(r)=\frac{C_{12}^{(e)}}{r^{12}}-\frac{C_6^{(e)}}{r^6}-\sigma\frac{C_3^\Omega}{r^3},
\end{equation}
with parameters determined by fits to experiments as $C_6^{(e)}=2810 \rm{a.u.}$, $C_{12}^{(e)}=1.862\times10^8 \rm{a.u.}$ and  $C_{3}^{\Omega=1}=-C_{3}^{\Omega=0}/2=0.09695 \rm{a.u.}$ for the $1_u$ and $0_u$ states respectively.   The Hund's case-(c) variables, however, are not good quantum numbers in the region of interest.  Coriolis forces mix nuclear rotation and electronic angular momentum and hyperfine interaction mixes this with nuclear spin \cite{tiesinga2005}.  As such, the only good quantum numbers are the total angular momentum and its projection which we denote $T, M_T$; parity is fixed here to be -1 for the $p$-wave collisions.  

\begin{figure}[floatfix]
\begin{tikzpicture}
\tikzstyle{every node}=[font=\normalsize]
  \node (pic) at (0, 0) {\includegraphics[width=7.7cm]{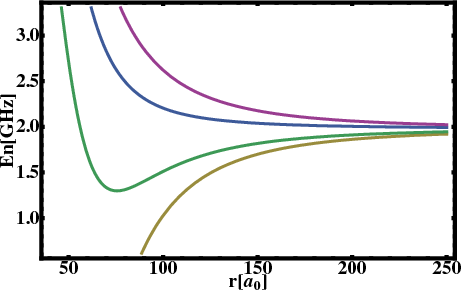}};
	\node at (2.4, 0.6) {\small $^1S_0+{}^3P_1(f_2=3/2)$};
%    \path[sloped] (-4.1, -2.0) -- node {$\bm\psi_{\bm n \bm , \bm \epsilon\bm(\bm T \bm , \bm M_{\bm T}\bm )}\bm (\bm r\bm)$} (-4.1, +3);
\end{tikzpicture}
\caption{\label{fig:adiabpot} Adiabatic potentials for the four channels with $T=3$ that asymptote to the $^1S_0+{}^3P_1(f_2=3/2)$ channel. Since $M_T$ takes seven values, each channel is seven fold degenerate.}
\end{figure}

We are interested in the molecular bound states, or photoassociation resonances of these electronic potentials. Dipole selection rules break the resonances into two parity classes -- those accessible from $R$-even or $R$-odd ground states~\cite{enomoto2008, iris_2009}. Of particular interest are the PLR states arising from avoided crossings due to hyperfine mixing.  Figure \ref{fig:adiabpot} shows the adiabatic potentials with $T=3$ that asymptote to the $^1S_0+{}^3P_1(f_2=3/2)$ channel, where $f_2$ is the hyperfine quantum number of the excited state atom.  There exists one potential with its minimum at $\sim 75 a_0$ and a depth of 0.68 GHz.  This shallow potential nonetheless supports bound states that are well resolved and can be used for $p$-wave OFRs with suppressed three-body recombination.  

To determine the photoassociation resonances, we employ a multichannel integration of the Schr\"{o}dinger equation as discussed in~\cite{iris_2009}.  We consider first the case of no external magnetic fields.  The effective potential operator in the  $^1S_0+{}^3P_1$ channel, \Eref{Veff},  is written as a matrix expanded in the extended Hund's case-(e) basis $|\epsilon(T, M_T)\rangle \equiv |f_2, F, R, T, M_T\rangle$, where $\mathbf{F}=\mathbf{f}_1+\mathbf{f}_2$ and $\mathbf{T}=\mathbf{F}+\mathbf{R}$ \cite{tiesinga2005}.  Here $f_1=1/2$ is the spin of the ground-state atom, and $f_2=3/2, 1/2$ is the hyperfine spin of the excited-state atom. In the ground state $I=F=1$, $M_F=1$, $R=1$, and the total angular momentum takes the possible values $T_g=0,1,2$.  By dipole selection rules, in the excited channels the allowed values are therefore $T_e=0,1,2,3$. The effective excited potential matrix thus has 19 channels each of which are $2T_e+1$ fold degenerate, resulting in a total of 89 channels.   We denote the multichannel excited bound states as (neglecting the subscript $e$),
\begin{equation}\label{expand}
|n,T, M_T\rangle=\sum_{\epsilon(T, M_T)}{\psi_{n, \epsilon(T, M_T)}(r)|\epsilon(T, M_T)\rangle }.
\end{equation} 
In the binding energy range of $-1022$ MHz to $-3$ MHz, the system supports 2 bound states with $T=0$, 26 bound states with $T=1$, 15 bound states with $T=2$, and 23 bound states with $T=3$.

\begin{figure}[floatfix]
\begin{tikzpicture}
\tikzstyle{every node}=[font=\normalsize]
  \node (pic) at (0, 0) {\includegraphics[width=7.5cm]{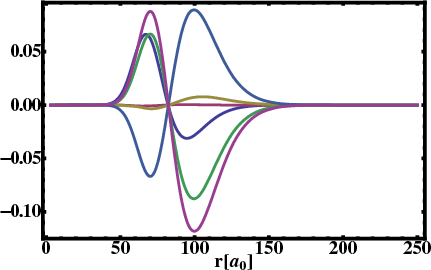}};
   \path[sloped] (-4.1, -2.0) -- node {$\psi_{\bm n \bm , \bm \epsilon\bm(\bm T \bm , \bm M_{\bm T}\bm )}(r)$} (-4.1, +3);
\end{tikzpicture}
\caption{\label{fig:PLR} Spinor components of the multichannel wave function of the PLR bound molecular state at -355 MHz and with $T=3$. Each curve corresponds to a component associated with one of the six  basis states, $| \epsilon(T, M_T) \rangle$, that contribute to this state. Since $M_T$ can assume seven distinct values, each wave function is 7-fold degenerate.}
\end{figure}

Of particular interest are the PLR states, denoted in Table \ref{vopt}(a).  Figure \ref{fig:PLR} shows an example of a multichannel spinor wave function of the PLR bound molecular state at $-355$ MHz with $T=3$. Each spinor component corresponds to one of the  six different $|\epsilon(T, M_T)\rangle$ channels, each of which are 7-fold degenerate.  Most of the amplitude of the wave function is supported between $50 a_0$ and $150 a_0$.  As such, the inner turning point is well removed from the chemical binding region and the outer turning point is sufficiently far out to allow for a large Franck-Condon factor in optical excitation.  These features are advantageous for application to OFRs.

 \subsection{In an external magnetic field}
We now consider the effect of an external magnetic field to allow for additional control on the system. With the $\mathbf{B}$-field defining the quantization axis and in the linear Zeeman regime, the perturbing potential is 
\begin{equation}\label{hb}
V_{\text{mag}}= \sum_{f_2, m_{f_2}} g_{f_2}\mu_B B \, m_{f_2} | f_2, m_{f_2} \rangle \langle f_2, m_{f_2} | ,
\end{equation} 
where $g_{f_2}$ is the Land\'{e} g-factor of the atomic hyperfine level.  
 This Hamiltonian breaks the rotational symmetry and generally couples an infinite hierarchy of states with different total angular momenta $T$. For the relatively weak magnetic fields that we consider here, we can employ perturbation theory.  We break the degeneracy of the states within a $T$-manifold and mix states with the same $M_T$ when the Zeeman shift is on the order of the vibrational spacing. For these weak magnetic fields, the value of $T$ at zero magnetic field still dominates and this will be used to label the states.   This is particularly true for the PLR states, where $T$ remains approximately a good quantum number for all fields we use in our calculation.
 %For the ground state, due to its vanishing electronic angular momentum,  the Zeeman Hamiltonian essentially vanishes.   
% Comment from Ivan:  I put this statement earlier when we define the ground state interactions.
 
To obtain the eigenenergies and eigenfunctions in the magnetic field, we diagonalize $V_{\text{mag}}$ expressed as a matrix in the basis of the bound states $|n,T, M_T\rangle$ within the energy range given in the discussion following Eq.\ (4).  The matrix elements are given by
\begin{widetext}
\begin{equation}
\langle n , T, M_T|V_{\text{mag}}|n', T', M'_{T}\rangle=
\sum_{\epsilon(T, M_T), \epsilon'(T', M_{T})}{\langle\epsilon(T, M_T)|V_{\text{mag}}|\epsilon'(T', M_{T})\rangle\int{\psi^*_{n, \epsilon(T, M_T)}(r)\psi_{n', \epsilon'(T', M_{T})}(r)}dr}\delta_{M_T,M'_T}.
\end{equation}
\end{widetext}
The term $\langle\epsilon(T, M_T)|V_{\text{mag}}|\epsilon'(T', M_{T})\rangle$ characterizes the coupling of the spin degrees of freedom and  the Franck-Condon overlap, $\int{\psi^*_{n, \epsilon(T, M_T)}(r)\psi_{n', \epsilon'(T', M_{T})}(r)}dr$, is the coupling of the radial wave functions.

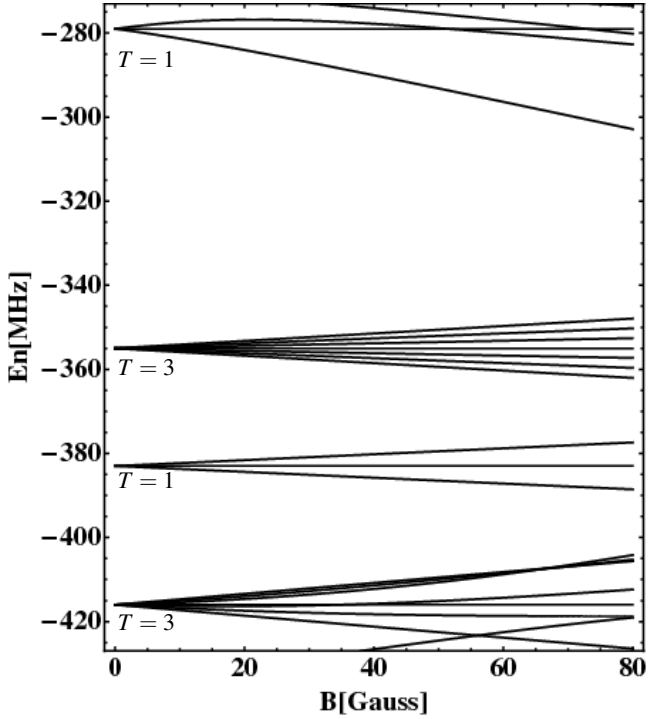
\begin{figure}
\begin{tikzpicture}
    \node (pic) at (0, 0) {\includegraphics[width=8.5cm]{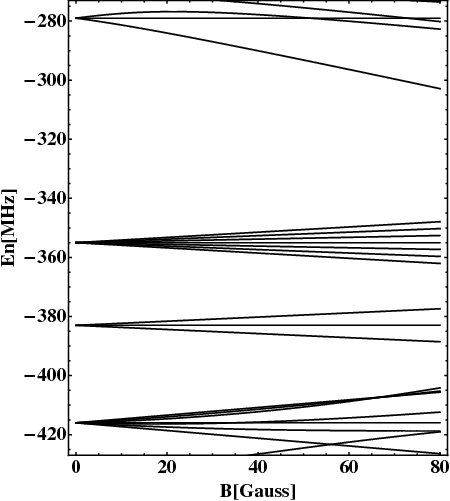}};
    \node at (-2.4, -0.1) {$T=3$};
    \node at (-2.4, 4.0) {$T=1$};
    \node at (-2.4, -1.6) {$T=1$};
    \node at (-2.4, -3.5) {$T=3$};
\end{tikzpicture}
\caption{\label{fig:envsB} Eigenspectrum of states in Table \ref{vopt}(b) (energy range: -427 MHz to -273 MHz) as a function of an applied magnetic field, $B$. For $B=0$ the eigenenergies and the quantum number $T$ correspond to the values in the left most column of Table \ref{vopt}(b).  We calculate the OFR associated with tuning near the $T=3$ PLR state, bound by $-355$ MHz, which splits into 7 magnetic sublevels in a linear Zeeman regime of the 80 Gauss plotted here}
\end{figure}

A part of the eigenspectrum, between -427 MHz and -273 MHz, is shown in Fig.~\ref{fig:envsB}.  The PLR state of interest, with binding of $355$ MHz and $T=3$, exhibits an approximately linear Zeeman splitting of its 7 magnetic sublevels over a range of 80 Gauss, as shown in the inset. Figure~\ref{fig:envsB} also shows that the  PLR state, with binding of $383$ MHz and $T=1$, also has an approximately linear Zeeman splitting of its 3 magnetic sublevels over the 80 Gauss range, while the remaining two states (-279 MHz and -416 MHz), which are not PLR, show nonlinear Zeeman shifts over that range of perturbation.

\section{The $p$-wave OFR}
To calculate the effect of the OFR on the $p$-wave scattering volume we turn to the theory of Bohn and Julienne~\cite{bohnandjulienne99}.  In that formalism the laser field is chosen detuned close to, but off-resonance from, a given photoassociation resonance. Only one bound state in a closed channel is assumed to contribute to the modification of the scattering volume.  In practice, the  laser field can couple to multiple excited bound states, and in the far-off-resonance limit, all will contribute.  How such multiple resonances interfere and affect the OFR is a subject of continued research.  Here, we will choose parameters for which one PLR bound state dominates and calculate its contribution to the OFR in both elastic and inelastic terms.

For a single bound state, the effect of the OFR on the S-matrix in the incoming $^1S_0 + {}^1S_0$ channel is
\begin{equation}\label{Smat}
 S=e^{2i \eta_0} \frac{2 \Delta-i\left(\Gamma -\gamma \right)}{2 \Delta+i\left(\Gamma +\gamma \right)}.
\end{equation}
$\gamma$ is the molecular natural linewidth, $\Delta$ is the detuning of the laser from the bound molecular state (including the light-shift of that level), $\eta_0$ is the background phase shift, and the stimulated linewidth is, 
\begin{equation}
\Gamma = \frac{\pi}{2} \left(\frac{I}{I_{sat}} \right) \gamma_A^2 f_{FC}.
\end{equation}
Here $I$ is the laser intensity, $I_{sat} = 0.13$ mW/cm$^2$ is the atomic saturation intensity for $^3P_1$, and $\gamma_A/2\pi = 182$ kHz is the atomic linewidth.  The Franck-Condon factor, with rotational corrections, is
\begin{equation}
f_{FC}=\frac{ |\langle n,T,M_T | \mathbf{d}\cdot \boldsymbol{\epsilon}_L | \psi_g (k_r) \rangle |^2}{2 d_A^2}
\end{equation}
expressed here as the ratio of the free-to-bound transition molecular dipole moment for laser polarization $\boldsymbol{\epsilon}_L$ to the atomic dipole momentum $d_A^2=(3c^3\gamma_A)/(4\omega^3)$.  In the Wigner-threshold regime $ |\psi_g (k_r) |^2 \propto k_r^3 $, where $k_r$ is the wave vector of the relative coordinate momentum at the scattering energy, and thus we define $\mathcal{V}_{\rm{opt}}$ as the ``optical volume" in analogy with the ``optical length" for $s$-wave OFRs,
\begin{equation}
\mathcal{V}_{\text{opt}}=\frac{\Gamma}{2 k_r^3 \gamma}.
\end{equation}
This is the parameter that defines the strength of the $p$-wave OFR.

\begin{table*}
{\Large(a)}
\begin{tabular}{|c|c|c|c|c|c|c|c|c|c|}
\hline
\multirow{2}{*}{Energy MHz} & \multicolumn{9}{c|}{$V_{\text{opt}}$ $a_0^3$}\\
\cline{2-10}
& \multicolumn{3}{|c|}{$m_R=-1$} &
\multicolumn{3}{|c|}{$m_R=0$} &
\multicolumn{3}{|c|}{$m_R=1$}\\
\cline{2-10}
 & $q=-1$ & $q=0$ & $q=1$ &$q=-1$ & $q=0$ & $q=1$ & $q=-1$ & $q=0$ & $q=1$\\
\hline
$-279 (T=1)$ & $2.13712\times10^6$ & $4.3758\times10^6$ & $3.61051\times 10^6$ & $396736$ &$36724$ & &$674875$ & &\\
\hline
$\mathbf{-355* (T=3)}$ & $75605$ & $113407$ & $75605$ & $113407$ &$302419$ & $378024$ &$75605$ &$378024$&$1.13407\times10^6$\\
\hline
$-383* (T=1)$ & $139596$ & $231235$ & $972724$ & $11501$ &$255429$ &  &$158527$ & &\\
\hline
$-416 (T=3)$ & $143913$ & $215869$ & $143913$ & $215869$ &$575652$ & $719565$ &$143913$ &$719565$&$2.15869\times10^6$\\
\hline
\end{tabular}

\vspace{0.7cm}

{\Large(b)}
\begin{tabular}{|c|c|c|c|c|c|c|}
\hline
\multirow{2}{*}{Energy MHz} & \multicolumn{3}{c|}{$V_{\text{opt}}$ $a_0^3$ ($B=0$ Gauss)}& \multicolumn{3}{c|}{$V_{\text{opt}}$ $a_0^3$ ($B=30$ Gauss)}\\
\cline{2-7}
 $B=0$& $m_R=-1$ & $m_R=0$ & $m_R=1$ &$m_R=-1$ & $m_R=0$ & $m_R=1$ \\
\hline
$-279 (T=1)$ & $3.61051\times 10^6$ & & & 844623 & &\\
\hline
$\mathbf{-355* (T=3)}$ & $75605$  & $378024$ &$1.13407\times10^6$& $76339 $&$379073$& $1.13089\times 10^6$\\
\hline
$-383* (T=1)$ & $972724$ &  & &973575& &\\
\hline
$-416 (T=3)$ & $143913$ & $719565$ &$2.15869\times10^6$ & $26884$ & $478907$& $2.14338 \times 10^6$\\
\hline
\end{tabular}
\caption{\label{vopt} $p$-wave optical volumes ($\mathcal{V}_{\text{opt}}$) for the coupling of all possible initial states to four of the bound molecular states of the excited potential (the energies of these states and their $T$ value are shown in the first column ) with (a) different polarizations, (b) polarization $q=1$. Blank entries indicate that the particular combination of initial state, polarization and final state is forbidden. A * indicates that the particular state is a PLR state.  The PLR state bound at $-355$ MHz, denoted in bold face, is used for the OFR calculation presented here.}
\end{table*}

% \caption{\label{Bvopt} $p$-wave optical volumes ($V_{\text{opt}}$) for the coupling of all possible initial states to four of the bound molecular states of the excited potential (energies of these states for $B=0$ are shown in the first column ) with $q=1$. Blanks indicate that the particular combination of initial state, polarization and final state is not possible. A * indicates that the particular state is a PLR state.

Selection rules dictate the allowed transitions that are accessible for an OFR.  For atoms with spin-polarized nuclei scattering on the ground $^1S_0+{}^1S_0$ potential, $I=1$, $m_I=1$, $T_g=0,1,2$, and $M_{T_g}=m_R+1$, where $m_R = -1,0,1$ are the projections of the partial-wave angular momentum on the quantization axis. We can thus optically connect to excited $^1S_0+{}^3P_1$ bound molecules with $T_e=0,1,2,3$, and $M_{T_e} = M_{T_g}+q=m_R+1+q$, where $q$ denotes the projection of photon helicity $(\pi, \sigma_\pm)$.  Table \ref{vopt}(b) shows the values of $\mathcal{V}_{\text{opt}}$ for the coupling of the partial-wave projections $m_R$ to four of the excited bound molecular states using $\sigma_+$ polarized light at $B=0$ Gauss and $B=30$ Gauss.  A non-zero $\mathbf{B}$-field leads to a mixing between the different eigenstates with the same $M_{T_e}$, leading to a change in the Condon overlap of this state with the scattering wave function of the ground potential. For the PLR states we see that the $\mathcal{V}_{\text{opt}}$ is fairly constant, while for the other states the $\mathcal{V}_{\text{opt}}$ is significantly changed by the magnetic field. This is because the poor overlap of the PLR states with their neighboring non-PLR states suppresses mixing.  

% 
% \begin{table*}
% \begin{tabular}{|c|c|c|c|c|c|c|}
% \hline
% \multirow{2}{*}{Energy MHz} & \multicolumn{3}{c|}{$V_{\text{opt}}$ $a_0^3$ ($B=0$ Gauss)}& \multicolumn{3}{c|}{$V_{\text{opt}}$ $a_0^3$ ($B=30$ Gauss)}\\
% \cline{2-7}
%  $B=0$& $m_R=-1$ & $m_R=0$ & $m_R=1$ &$m_R=-1$ & $m_R=0$ & $m_R=1$ \\
% \hline
% $-279 (T=1)$ & $3.61051\times 10^6$ & & & 844623 & &\\
% \hline
% $\mathbf{-355* (T=3)}$ & $75605$  & $378024$ &$1.13407\times10^6$& $76339 $&$379073$& $1.13089\times 10^6$\\
% \hline
% $-383* (T=1)$ & $972724$ &  & &973575& &\\
% \hline
% $-416 (T=3)$ & $143913$ & $719565$ &$2.15869\times10^6$ & $26884$ & $478907$& $2.14338 \times 10^6$\\
% \hline
% \end{tabular}
% \caption{\label{Bvopt} $p$-wave optical volumes ($V_{\text{opt}}$) for the coupling of all possible initial states to four of the bound molecular states of the excited potential (energies of these states for $B=0$ are shown in the first column ) with $q=1$. Blanks indicate that the particular combination of initial state, polarization and final state is not possible. A * indicates that the particular state is a PLR state.}
% \end{table*}

With the scattering matrix in hand, the $p$-wave scattering volume is defined as $a_p^3 = -K/k_r^3$, where the $K$-matrix element is given by~\cite{taylor}
\begin{equation}\label{Kmat}
 K=i\frac{1-S}{1+S}=-\frac{\Gamma/2}{\Delta +i\gamma/2} ,
\end{equation}
 excluding the background phase shift.  The real and imaginary parts of the $p$-wave scattering volume are then
\begin{eqnarray}\label{scatt_vol}
\Re {(a_p^3)}=a_{bg}^3+\mathcal{V}_{\rm{opt}}\frac{\gamma\Delta }{\Delta^2 +\frac{\gamma^2}{4}},\\
 \Im{(a_p^3)} = -\frac{\mathcal{V}_{\rm{opt}}}{2} \frac{\gamma^2}{\Delta^2 +\frac{\gamma^2}{4}}.
\end{eqnarray}
where the additional background contribution was added.  We obtain the background phase shift $\eta_{0}$ by numerical integration of the $p$-wave scattering state in the Wigner threshold regime and fit to the asymptotic wave function.  We find the background scattering volume to be $a_{bg}^3 =-406446$ a.u.  

The real and imaginary parts of the scattering volume govern the strengths of the elastic and inelastic collisions respectively.  In principle, one can increase the ratio of good to bad collisions solely by increasing the detuning.  In practice, this is limited by the available intensity that is required to ensure a sufficiently strong interaction.  Moreover, our model is restricted to an OFR via a single molecular bound state, and for self-consistency, we require a sufficiently small detuning so that only one photoassociation resonance dominates the process.  For these reasons, we must choose a state in a sufficiently sparse region of the density of states so that when the laser is detuned closest to this state, even for detunings large enough to avoid spontaneous scattering, the single resonance model is valid.

We thus seek a PLR state that we can address with high resolution and with a sufficient optical volume to induce a strong OFR.  Firstly, as we are considering spin-polarized fermions, we can ignore the nearby spectrum of $s$-wave photoassociation resonances and concentrate only on the bound states connected to $p$-waves.  Secondly, by employing dipole selection rules, we can reduce the number of allowed transitions and reduce the density of states.  Using a magnetic field and polarized light, the interaction strength for scattering in states of the ground potential with a particular $m_R$ value can be selectively enhanced while suppressing the interaction strength for scattering in states with other $m_R$ values. For example, the state with $m_R=1$, corresponding to the stretched state, $T_g=2, M_{T_g}=2$,  couples with $\sigma_+$ polarized light  only to a $T_e=3, M_{T_e}=3$ molecular bound state. Transition to states with other values of $T_e$ are forbidden.  Of course, the ground state can not be prepared in a state with a given $m_R$, but in the presence of a magnetic field, differences in detuning and optical volumes can suppress other transitions.

Given these observations, the PLR at $-355$ MHz is promising for application to a $p$-wave OFR.  This is a $T_e=3$ state which connects only to a $T_g =2$ ground-state. In the presence of a magnetic field and with $\sigma_+$ polarized light, we can address the ground $M_{T_g} = 2 \rightarrow M_{T_e} = 3$ transition and make this the dominant resonance (see Fig.~\ref{fig:energy_levels}).  The neighboring bound states are $T_e =1$ (see Fig. 3) and inaccessible with this polarization from the $T_g=2, M_{T_g}=2$ ground state.  In addition, the $\mathcal{V}_{\text{opt}}$ for $m_R=1$ is substantially larger than $\mathcal{V}_{\text{opt}}$ for the other sublevels indicating that the OFR is strongest for the $ M_{T_g}=2$ state. This leads to a further enhancement of the interaction strength for the $M_{T_g} = 2 \rightarrow M_{T_e} = 3$ transition.

Figure~\ref{fig:energy_levels} shows a possible configuration for inducing the OFR.  In a 30 Gauss magnetic field, and  detuning $\Delta = -3\, \rm{MHz}$ below the resonance at $-355$ MHz, we dominantly couple the $T_g=2, M_{T_g} = 2 \rightarrow T_e=3, M_{T_e} = 3$ transition.  Using \Eref{scatt_vol} we calculate the real-part of the scattering volume arising from the OFR to be $\Re(a_p^3) = -1.44 \times 10^5  (\rm{W/cm}^2)^{-1}$.  The imaginary part is reduced by the factor $\gamma/2 \Delta = 0.057$.  The effect of this spontaneous emission will depend on the application at hand.   Coupling to other transitions, $M_{T_g} = 0,1 \rightarrow  M_{T_e} = 1,2$ are reduced to $\Re(a_p^3) = -4.36 \times 10^4 (\rm{W/cm}^2)^{-1}$ and $\Re(a_p^3) = -1.40 \times 10^4 (\rm{W/cm}^2)^{-1}$ respectively.  In addition, off-resonant coupling of  $T_g=2, M_{T_g} = 0$ to neighboring  $T_e=1, M_{T_e} = 1$ is highly suppressed at this detuning.

\section{The three-color Fermi-Hubbard model}
It will be extremely challenging to observe $p$-wave superfluidity in a dilute gas, even with the use of an OFR, given the ultra-low temperatures required.  Nonetheless, the ability to control $p$-wave interactions can potentially lead to a rich variety of many-body phenomena, particularly if an optical lattice confining potential is included. We propose here how the combination of such tools can be used to explore a toy model of fermionic color superfluidity with three colors.  Such models have been considered before~\cite{rapp2007} where the {\em internal} degrees of freedom served as the three ``colors''.  For the case of fermions, this is not a natural realization since the number of internal states will always be an even number.  An alternative is to employ the {\em external} degrees of freedom associated with the three  spatial orbitals of the first excited ``$p$-band'' of an optical lattice.  Such colors have been considered for bosons, mediated by s-wave interactions.  We consider here a model for spin polarized fermions, mediated by $p$-wave interactions.  
 
Following~\cite{pxpyhubbard}, the multicolor field operator for spinless (i.e. polarized) fermions in the first excited $p$-band is written in the Wannier basis as
\begin{equation}
\psi(\mathbf{x})=\sum_{i,\alpha} c_{i,\alpha} \phi_{\alpha} (\mathbf{x}- \mathbf{R}_i ) ,
\end{equation}
where $\phi_{\alpha} ( \mathbf{x} )$ is a p-orbital with $\alpha = x,y,z$, and $c_{i,\alpha}$ is the fermionic annihilation operator for that orbital at the $i^{th}$ lattice site.  We consider lattices of sufficient depth $V_0$ that the tight-binding approximation is valid.  We restrict the dynamics to a single $p$-band, which can be metastable, as seen in recent experiments where bosons remained in the first excited band of an optical lattice for about a hundred times the tunneling time scale~\cite{pband_boson_raman, pband_boson_doublewell}. We expect a similar metastability for fermions.  In addition, we assume sufficiently deep lattices such that the tunneling coefficient for a particle in the state $\alpha$ in negligible along the direction $\alpha'$ for $\alpha' \neq \alpha$.  Moreover, we take the wells to be spherically symmetric.  The Hamiltonian then takes the Fermi-Hubbard form for the three colors in a single band,
 \begin{equation}
 \label{hubbard}
 H=-J \sum_{\langle i,j\rangle_\alpha,\alpha} {c_{i,\alpha}^\dagger c_{j,\alpha}} +
 \sum\limits_{i, m_R, \alpha \beta, \alpha' \beta'} {c_{i,\alpha'}^\dagger c_{i,\beta'}^\dagger c_{i,\alpha} c_{i,\beta} \, V^{m_R}_{\alpha' \beta',\alpha \beta}},
\end{equation}
where $J$ is the tunneling coefficient along any direction $\alpha$, $\langle i, j \rangle_\alpha$ indicates that $i$ and $j$ are nearest neighbors along $\alpha$, and $V^{m_R}_{\alpha' \beta',\alpha \beta}$ is the interaction matrix element for two atoms at the same site starting in orbitals $\alpha,\beta$ and scattering to $\alpha',\beta'$ via $p$-wave collisions of symmetry $m_R$.   The coupling matrix is
\begin{equation}
V^{m_R}_{\alpha' \beta',\alpha \beta} = \int \phi^*_{\alpha'} (\mathbf{x}_1) \phi^*_{\beta'} (\mathbf{x}_2) V^{m_R}_p (\mathbf{x}_1-\mathbf{x}_2) \phi_{\alpha} (\mathbf{x}_1) \phi_{\beta} (\mathbf{x}_2),
\end{equation}
where $V^{m_R}_p (\mathbf{x}_1-\mathbf{x}_2)$ is the two-body interaction potential for $p$-wave scattering.  This can be treated through a pseudopotential on a delta-shell \cite{stock2005}
\begin{equation}
V^{m_R}_p(\mathbf{r}) = \lim_{s \rightarrow 0}\frac{3  \Re(a_p^3)}{4\mu}\, Y_{1,m_R}(\theta_r,\phi_r) \, \frac{\delta(r-s)}{s^3} \frac{\partial^3}{\partial r^3}(r^2 \, \, \, ).
\end{equation}
In order to calculate the interaction matrix we transform the Wannier states from the Cartesian orbitals to spherically symmetric 3D harmonic oscillator orbitals, and to center-of-mass and relative coordinates of the two particles, specified by the projections of angular momentum, $M_R$ and $m_R$, respectively.  The matrix then takes the form
\begin{equation}
V^{m_R}_{\alpha' \beta',\alpha \beta} =  \sum\limits_{M_R} \langle \alpha' \beta' | m_r M_R \rangle U^{m_R}\langle m_r M_R |  \alpha \beta \rangle,
\end{equation}
where $\langle m_r M_R |  \alpha \beta \rangle$ is the angular part of the change-of-basis matrix, and $ U^{m_R}$ is the interaction strength coming from the radial integral of the interaction potential expressed in the relative coordinate, proportional to the real-part of the $p$-wave scattering volume.

Like the model studied in \cite{rapp2007}, the Fermi-Hubbard Hamiltonian \Eref{hubbard} has three colors, but differs in two important ways.  Firstly, it allows for anisotropic interactions as considered in \cite{miyatake2009}. In addition, we allow for couplings between different incoming and outgoing orbitals, $\alpha \neq \alpha', \beta \neq \beta'$, as studied for bosons, \cite{pxpyhubbard}.  Most importantly, unlike any model previously considered, the control provided by the OFR allows for the possibility to manipulate the strength of interactions in a manner that depends on the fermionic colors. We expect such  control could be used to explore a variety of phenomena such as the trionic phase and color superfluids discussed in Ref.~\cite{rapp2007}.  We leave the details of the many-body analysis for future work.

\begin{figure}
      \begin{tikzpicture}[scale=1.2]
     \filldraw[green, ultra thick] (-1,0) circle (0.1cm);
     \filldraw[green, ultra thick] (0.2,0) circle (0.1cm);
     \filldraw[green, ultra thick] (1.4,0) circle (0.1cm);
     \draw[ultra thick] (-1.3,0) -- (-0.7,0);
     \draw[ultra thick] (-0.1,0) -- (0.5,0);
     \draw[ultra thick] (1.1,0) -- (1.7,0);

%      \draw[ultra thick] (-4.9,6.3) -- (-4.3,6.3);
%      \draw[ultra thick] (-3.7,5.8) -- (-3.1,5.8);
%      \draw[ultra thick] (-2.5,5.3) -- (-1.9,5.3);
%      \draw[ultra thick] (-1.3,4.8) -- (-0.7,4.8);
     \draw[ultra thick] (-0.1,4.3) -- (0.5,4.3);
     \draw[ultra thick] (1.1,3.8) -- (1.7,3.8);
     \draw[ultra thick] (2.3,3.3) -- (2.9,3.3);
     \draw[->, red, ultra thick] (-1,0) -- (0.0,2.7);
     \draw[->, red, ultra thick] (0.2,0) -- (1.1,2.7);
     \draw[->, red, ultra thick] (1.4,0) -- (2.4,2.7);

%      \draw[ultra thick] (0.2,0) -- (0.2,4);

     \draw[dashed, blue, ultra thick] (-1,2.7) -- (2.9,2.7);

	\node at (-1, -0.3) {{$m_R=-1$}};
	\node at (0.2, -0.3) {{$m_R=0$}};
	\node at (1.4, -0.3) {{$m_R=1$}};

	\node at (-1, -0.6) {{$M_{T_g}=0$}};
	\node at (0.2, -0.6) {{$M_{T_g}=1$}};
	\node at (1.4, -0.6) {{$M_{T_g}=2$}};

	\node at (0.2, 4.5) {{{${M_{T_e}=1}$}}};
	\node at (1.4, 4.0) {{{${M_{T_e}=2}$}}};
	\node at (2.6, 3.5) {{{${M_{T_e}=3}$}}};

	\node at (0.6, 3.3) {{$\Re{(a_p^3)}=-7.23\times 10^6$ $a0^3$}};
	\node at (-0.6, 3.8) {{$\Re{(a_p^3)}=-2.18\times 10^6$ $a0^3$}};
	\node at (-1.8, 4.3) {{$\Re{(a_p^3)}=-6.97\times 10^5$ $a0^3$}};

	\node at (0.4, 2.4) {\color{red}{$q=1$}};
% 	\node at (2.6, 6.3) {\color{pink}{$B=30$ G,}};
% 	\node at (2.6, 5.6) {\color{pink}{$\Delta=-3$ MHz,}};
% 	\node at (2.6, 4.9) {\color{pink}{$I=10$ $W/{cm}^2$}};

	\draw [dgreen,decorate,decoration={brace,amplitude=5pt},ultra thick]
	      (2.9,3.3)  -- (2.9,2.7);

	\node at (3.2, 3) {{$\Delta = 3$ MHz}};
% 	\draw [dgreen,decorate,decoration={brace,amplitude=5pt},ultra thick]
% 	      (2.0,3.3)  -- (2.0,3.8);
% 
% 	\node at (1.2, 3.5) {{$0.89$ MHz}};
% 
% 	\draw [dgreen,decorate,decoration={brace,amplitude=5pt},ultra thick]
% 	      (1.0,3.8)  -- (1.0,4.3);
% 
% 	\node at (0.2, 4.1) {{$0.89$ MHz}};
% 	\draw [dgreen,decorate,decoration={brace,amplitude=5pt},ultra thick]
% 	      (1.7,4.8)  -- (1.7,5.3);

% 	\node at (0.8, 5.05) {\color{dgreen}\scriptsize{$0.88$MHz}};

    \end{tikzpicture}
\caption{\label{fig:energy_levels} OFR using $\sigma^+$ polarized light to couple the scattering state of the ground potential with the three different projections of $p$-wave angular momentum, $m_R$, to the excited PLR bound state with the different total projection $M_{T_e}$.  The figure shows only those  states permitted by selection rules.  Denoted are OFR values of the real part of the scattering volume, $\Re(a_p^3)$, for each of the three transitions, for an intensity of 50 W/cm$^2$.}
\end{figure}
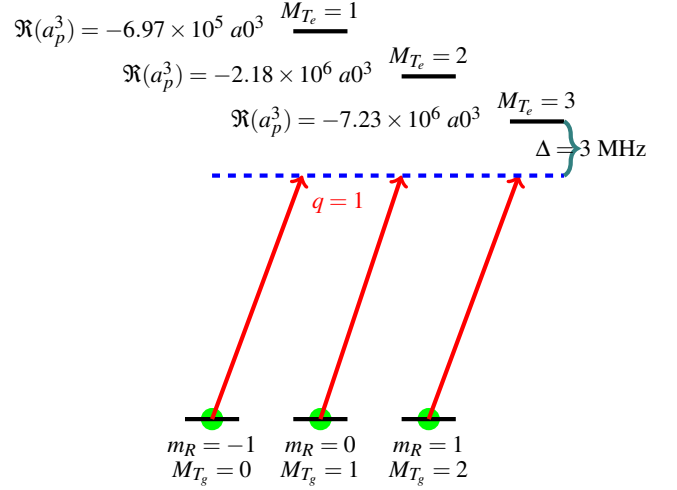

To evaluate the potential for this system to lead to quantum critical behavior, we give here a rough back-of-the-envelope estimate. We expect that interesting many-body physics will be accessible when the ratio between the kinetic and interaction energies in the system is of order one~\cite{u6j}, i. e., 
$ U_{m_R}\gtrsim 6J $,
for some given choice of $m_R$. Choosing the lattice depth along any direction to be $V_0=18 E_r$, where $E_r$ is the recoil energy, we find $J= 0.16 E_r$ in the first excited band.  For these parameters, it follows that phase transitions occur near
\begin{equation}\label{criterion}
 |\Re{(a_p^3)}|\gtrsim 7 \times 10^6 \text{a.u.}
\end{equation}
Typically, the $p$-wave scattering volume arising from the background phase shift is very small and the model in \Eref{hubbard} does not result in quantum phase transitions. However, using an OFR, $\Re {(a_p^3)}$ can be tuned to larger values.  Moreover, through selection rules we can control specific $m_R$-couplings that correlate with interactions of specific colors.  Figure~\ref{fig:energy_levels} outlines one possible scheme. The different $M_{T_g}$ levels are coupled to specific $M_{T_e}$ levels in the excited PLR state, shown in Fig.~\ref{fig:envsB}, using $\sigma^+$ polarized light.  The Zeeman splitting between the different $M_{T_e}$ levels of the excited state is approximately $0.89$ MHz at $B=30$ Gauss. The figure indicates the values of $\Re{(a_p^3)}$, calculated for the couplings between the different states for a laser of intensity $I=50 \text{W/cm}^2$ and a detuning $\Delta =3$ MHz below the $M_{T_e}=3$ state. For this magnetic field, laser intensity, polarization, and detuning the atoms, scattering in the $m_R=1$ state will experience a $p$-wave scattering volume of $\Re(a^3_p) = - 7.24 \times 10^6$ a.u.,  satisfying the criterion in \Eref{criterion}.  With such control, we expect one can observe novel quantum critical behavior in the fermionic superfluid.

\section{Summary and outlook}
We have studied a highly controllable system of spin-polarized ${}^{171}$Yb atoms undergoing $p$-wave collisions as modified by an optical Feshbach resonance (OFR).  By tuning near an electronically excited purely-long-range (PLR) bound state in the $^1S_0 + {}^3P_1$ channel, we expect to suppress three-body recombination losses that typify magnetically induced $p$-wave Feshbach resonances in the ground electronic manifold.  We used a multichannel integration of the Schr\"{o}dinger equation to determine the photoassociation resonances and the eigenfunctions including perturbing magnetic fields.  With these, we calculated the real and imaginary part of the ``scattering volume'' associated with the $p$-wave scattering phase shift and loss rate for choices of magnetic fields and OFR polarized laser fields.  

Because the ${}^3P_1$ state has a relatively large linewidth as compared to the other intercombination lines, the demands on precision control of laser detuning are moderate.  On the other hand, this larger linewidth implies a limitation on the strength of the real-part of the Feshbach resonance before inelastic scattering can no longer be neglected.  For these reasons we expect that even with an OFR, one will not be able to achieve $p$-wave superfluidity, or a BEC-BCS crossover, analogous to that seen for s-wave pairing.  Nonetheless, the degree of control afforded by the OFR could open the door to explorations of novel quantum critical behavior in the many-body system.  

We began such an exploration, considering a new model of  three-color fermonic superfluidity.  Here the three colors correspond to the three spatial orbitals of spinless (i.e. polarized) fermions in the first excited $p$-band of an optical lattice.  Based on this toy model, we calculated the parameters of a Hubbard model including nearest neighbor hopping and on-site interaction between two fermions in different orbitals via $p$-wave collisions.  Through careful choice of magnetic field, laser polarization, and detuning, we find conditions under which tunneling and interaction energy scales are comparable.  For such operating conditions, we expect quantum phase transitions are possible.  A full many-body exploration of the phase diagram is left for future analysis.

We thank Maciej Lewenstein, Pietro Massignan, and Philipp Hauke for helpful discussions, particularly about the application of our model to the Fermi-Hubbard Hamiltonian. KG and IHD acknowledge support from the Office of Naval Research Grant No. N00014-03-1-0508  and the Center for Quantum Information and Control (CQuIC) via the National Science Foundation Grant PHY-0969997.

\bibliography{v22}

\end{document}